\documentclass[dvips,aps,pra,preprint,showkeys,longbibliography]{revtex4-1}

\usepackage{dcolumn}			
\usepackage{bm}					

\usepackage{placeins}			
\usepackage[dvips]{graphicx} 	
\usepackage{pslatex}	     	
\usepackage{amsmath,amssymb,amsfonts}
\usepackage{esint}
\usepackage{nicefrac}
\usepackage{xstring}  			
\usepackage{stackengine}		
\usepackage{siunitx}
\usepackage{scrextend} 			

\newcommand{\Tens}[2]{  
\IfEqCase{#2}{%
	{1}{\underline{#1}}
	{2}{\underline{\underline{#1}}}
	{3}{\underline{\underline{\underline{#1}}}}
	{4}{\underline{\underline{\underline{\underline{#1}}}}}
	}[\PackageError{Tensor}{Undefined option to Tensor: #2}{}]%
}

\newcommand{\Tensm}[1]{
\IfEqCase{#1}{
	{2}{\mathrel{\Shortstack{{.} {.}}}}
	{3}{\mathrel{\Shortstack{{.} {.} {.}}}}
	}[\PackageError{Tensm}{Undefined option to Tensm: #1}{}]%
}

\newcommand{\jj}{j}

\newcommand{\conj}[1]{\bar{#1}}
\newcommand{\abs}[1]{\left| #1 \right|}
\newcommand{\ee}{\varepsilon}

\newcommand{\Exp}[1]{\mathrm{e}^{#1}}

\newcommand{\mm}{\vec{\mathbf{\mu}}}

\begin{document}
\preprint{-}

\title{Multiferroic Micro-Motors With Deterministic Single Input Control}

\author{John P. \surname{Domann}}
\email[]{jpdomann@vt.edu}
\affiliation{Department of Biomedical Engineering and Mechanics, Virginia 
Polytechnic and State University}

\author{Cai \surname{Chen}}
\affiliation{Department of Mechanical and Aerospace Engineering, University of 
California, Los Angeles}

\author{Abdon E. \surname{Sepulveda}}
\affiliation{Department of Mechanical and Aerospace Engineering, University of 
California, Los Angeles}

\author{Rob N. \surname{Candler}}
\affiliation{Department of Electrical and Computer Engineering, University of 
California, 
Los Angeles}
\affiliation{Department of Mechanical and Aerospace Engineering, University of 
California, Los Angeles}
\affiliation{California NanoSystems Institute, Los Angeles}

\author{Greg P. \surname{Carman}}
\affiliation{Department of Mechanical and Aerospace Engineering, University of 
California, Los Angeles}

\date{\today}


\begin{abstract}
\textbf{Abstract:}\par
This paper describes a method for achieving continuous deterministic  
360$^{\circ}$ magnetic moment rotations in single domain magnetoelastic discs, 
and examines the performance bounds for a mechanically lossless multiferroic 
bead-on-a-disc motor based on dipole coupling these discs to small magnetic 
nanobeads. The continuous magnetic rotations are attained by controlling the 
relative orientation of a four-fold anisotropy (e.g., cubic magnetocrystalline 
anisotropy) with respect to the two-fold magnetoelastic anisotropy. This 
approach produces continuous rotations from the quasi-static regime up through 
operational frequencies of several GHz. Driving strains of only $\approx$90 to 
180 ppm are required for operation of motors using existing materials. The 
large operational frequencies and small sizes, with lateral dimensions of 
$\approx$100s of nanometers, produce large power densities for the rotary 
bead-on-a-disc motor, and a newly proposed linear variant, in a size range 
where power dense alternative technologies do not currently exist. 
\end{abstract}

\maketitle 

\section{Background / Literature Review}	\label{sec: Background}
In his 1842 lecture \textit{On a new Class of Magnetic Forces} James Joule 
describes how following the suggestion of an "ingeneous gentleman," he 
initiated the first ever measurements of magnetoelasticity.\cite{Joule1842, 
*Joule2011} From the outset, Joule was focused on determining not just if a bar 
of iron would elongate in the presence of a magnetic field, but if "power could 
be advantageously employed for the movement of machinery." In other words, 
Joule was interested in making magnetoelastic motors. However, after 
measuring a magnetic field induced strain of only $ \ee\approx1.4$ 
ppm, Joule concluded that "With regard to the application of the new force to 
the movement of machinery, I have nothing favourable to advance." 

\begin{figure*}[hpbt!]
	\includegraphics[width=17cm]{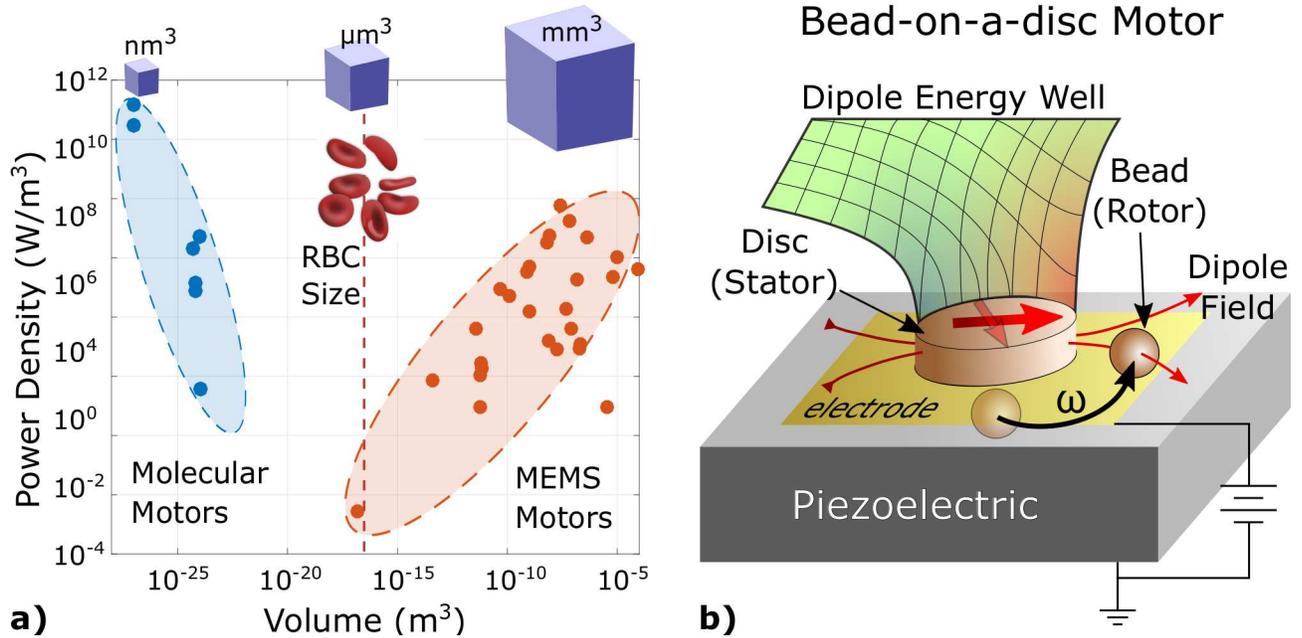}
	\protect\caption{\label{fig:motor power density scaling}a) Power density of 
	micro and nanoscale motor technologies, indicating no power dense 
	technologies exist at the \SIlist{}{\micro\meter^3} size 
	scale.\cite{Balzani2006, Lussis2011,Regan2005, Kang2009, Fuller2007, 
	Junge2009, Paenke2001, Liu2010, Kim2014, Tai1989, Wallrabe1992, Tai1989a, 	
	Liu2010a, Jacobsen1989,Ghalichechian2008, Livermore2004, Glickman2011, 	
	Chang2014, Achotte2006, Takemura2007, Kanda2006, Watson2009, Mashimo2015, 	
	Flynn1998, Sashida1993, Morita2000, Cagatay2004, Satonobu2000, Suzuki2002,	
	Watson2009a, DeVolder2011, Buetefisch2002, Konishi2001, DeVolder2008, 	
	Gebhard1997} Shown for scale are several red blood cells (RBCs). 
	b) The proposed multiferroic bead-on-a-disc motor uses dipole coupling to 
	drag a small magnetic bead around a stationary magnetic disc with rotating 
	magnetic moment.}
\end{figure*}

Of course, in that very same lecture Joule was doubtful that electromagnetic 
motors had any future use, showing a slight lack of foresight in an 
otherwise exemplary career. While many different types of motors have been 
created, macro-scale combustion and electromagnetic motors are the most 
widely used. However, poor scaling laws rapidly degrade the performance of 
these traditional motor technologies at the microscale, and new approaches need 
to be considered. Figure \ref{fig:motor power density scaling}a highlights the 
available power density for a variety of different micro and nanoscale motor 
technologies. For reference, the combustion engine of a 2017 Ford Mustang can 
generate around 400 horsepower (\SIlist{300}{\kilo\watt}) and is sized on the 
order of \SIlist{1}{\meter\cubed}, generating a power density of  
\SIlist{300}{\kilo\watt\per\meter\cubed} (i.e., moderate to high power density 
but at significantly larger sizes than shown on Figure \ref{fig:motor power 
density scaling}a). 

The larger sizes on Figure \ref{fig:motor power density scaling}a are  
representative of MEMS motors with volumes $\approx$ 
\SIlist{1}{\milli\meter\cubed}. The data points come from numerous different 
technological approaches, including electrostatic,\cite{Tai1989, Wallrabe1992, 
Tai1989a, Liu2010a, Jacobsen1989, Ghalichechian2008, Livermore2004} 
magnetostatic,\cite{Glickman2011, Chang2014} electromagnetic,\cite{Achotte2006} 
electrohydrodynamic,\cite{Takemura2007} piezoelectric,\cite{Kanda2006, 
Watson2009, Mashimo2015, Flynn1998, Sashida1993, Morita2000, Cagatay2004, 
Satonobu2000, Suzuki2002, Watson2009a} pneumatic,\cite{DeVolder2011, 
Buetefisch2002,Konishi2001, DeVolder2008, Gebhard1997} and 
plasmonic\cite{Liu2010, Kim2014} technology bases. These motors can 
advantageously be controlled with conventional electronic systems; however the 
power density of existing MEMS approaches clearly doesn't scale well to sizes 
significantly smaller than a cubic millimeter. 
	
At volumes near \SIlist{1}{\nano\meter^3}, molecular motors like rotaxane are 
extremely power dense. These molecules generate forces near 
\SIlist{30}{\pico\newton}, with strokes of \SIlist{5}{\nano\meter}, at rates of 
up to \SIlist{3}{\kilo\hertz}. This produces power densities on the order 
of 100 \si{\giga\watt\per\meter\cubed}.\cite{Balzani2006, Lussis2011} In 
addition to rotaxane, molecular motors based on nanocrystal 
actuators,\cite{Regan2005, Kang2009} and ATP synthase\cite{Junge2009, 
Paenke2001} have been created. However, molecular motors are exceedingly 
difficult to locally control, leading to poor array / scaling properties.

Figure \ref{fig:motor power density scaling}a illustrates that there 
are currently no power dense motor technologies at the \si{\micro\meter\cubed} 
size scale. A \si{\micro\meter\cubed} motor would enable localized control on 
the same size scale as biological cells, facilitating diagnostic and 
therapeutic applications in addition to opening new avenues for fundamental 
research. As an example, Di Carlo et al. have recently demonstrated magnetic 
cell sorting techniques near the microscale, and improved control could enable 
the study of individual cellular components instead of relying on traditional 
cell lysis based techniques.\cite{Murray2016}

Multiferroic heterostructures have recently been proposed as highly power dense 
\si{\micro\meter\cubed} motors.\cite{Sohn2015} These motors manipulate 
large magnetic forces by controlling the dipole fields of single 
domain magnetic heterostructures or domain walls with strain. Multiferroic 
control is energy efficient\cite{DSouza2016,DSouza2012} even at high frequency 
operation.\cite{Labanowski2017, Chen2017} These effects should combine to 
produce large power densities. Figure \ref{fig:motor power density scaling}b 
shows a hypothetical bead-on-a-disk motor,\cite{Sohn2015} where the magnetic 
moment of the large disk rotates (the disk itself is stationary), and drags 
around the smaller bead through dipole coupling. The B-field generated by the 
large disk applies a torque on the magnetic bead to keep it aligned 
with the B-field, while $ \nabla B $ applies a force on the bead and 
causes it to rotate around the large disk. 

A key impediment to the creation of multiferroic motors has been achieving 
deterministic control of continuous $ 360^{\circ} $ magnetic rotations. Initial 
studies on magnetoelastic control used single electrodes to cause $ 90^{\circ} 
$ non-deterministic rotations in ellipse shaped 
structures.\cite{Wu2011,Hockel2013,Sohn2015} However, the use of a single 
electrode and elliptical shape is restricted to a maximum rotation of $ 
90^{\circ} $, since it combines two uniaxial (i.e., two-fold) anisotropies (see 
the Supplementary Information for a mathematical explanation). This restriction 
motivated subsequent studies with multiple electrodes that generated a rotating 
biaxial plane strain state.\cite{Cui2013, Cui2015} Several modeling efforts 
have looked at this challenge, and observed that the use of 4 or more 
electrodes can deterministically control rotations.\cite{Liang2014, Liang2015} 
The multi-electrode approach has resulted in experimental observation of 
deterministic rotations in increments of $ 45^{\circ} $ confirmed up to $ 
180^{\circ} $.\cite{Sohn2017} Recent work has also shown dynamic effects can 
lead to $ 180^{\circ} $ rotations (i.e., ballistic switching with 
PMA).\cite{Wang2017a} However, fabricating motor arrays with multiple 
electrodes is undesirable, and ballistic switching requires precise timing with 
very narrow-band performance limiting variable frequency use.

A path to single electrode deterministic control with broadband operation can 
be found in three different studies analyzing the effects of rotating the 
relative orientation of the magnetoelastic anisotropy with respect to the shape 
of nanomagnetic structures. This approach has led to model based predictions 
of deterministic rotations in shapes like four leaf clovers,\cite{Wang2014a} 
squares,\cite{Peng2015} cat eyes,\cite{Cui2017} and peanuts.\cite{Cui2017}   
However, most of these shapes require the use of complex  fabrication 
techniques to resolve fine features, and are prone to the creation of pinning 
sites that have likely prevented rotation in previous work.\cite{Cui2015} To 
overcome these issues, the present study analyzes a single electrode control 
method for deterministic $ 360^{\circ} $ rotations, avoiding complex geometries 
or electrode patterns. A bead-on-a-disk motor is then analyzed and predictions 
provided for a mechanically lossless motor's performance (i.e., only magnetic 
and inertial forces are considered in the limit of zero fluidic drag). 

\section{Model}	\label{sec: Motor Model}
While previous studies have combined magnetoelasticity with an additional 
two-fold anisotropy (i.e., elliptical shapes), this work combines a four-fold 
anisotropy (e.g., cubic MCA) with the magnetoelastic anisotropy. This 
work demonstrates the key to generating deterministic $ 360^{\circ} $ rotations 
is to rotate the principal orientation of the anisotropies with respect to one 
another. In the following section the static and dynamic rotational 
characteristics of the stator element (large magnetic disk) with rotated 
anisotropies will be analyzed, then a second section analyzing the upper 
performance bounds of a bead-on-a-disk motor will be presented.

\subsection{Quasi-Static Energy Analysis}
\begin{figure*}[hptb!]
	\includegraphics[width=1.0\linewidth]{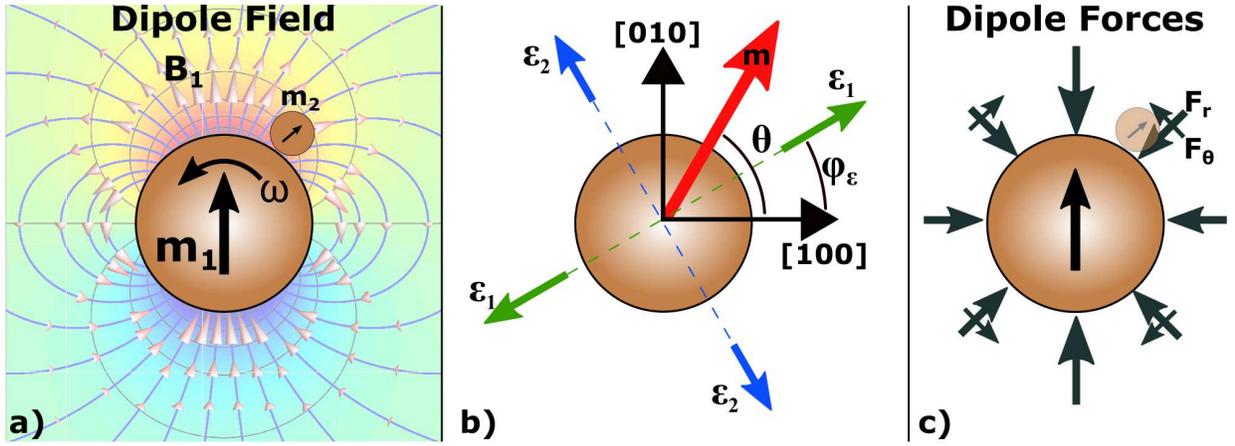}
	\caption{\label{fig:dipole_motor} a) Bead-on-a-disk motor with surrounding 
	magnetic dipole field. The large disk (stator) remains stationary, but its 
	magnetic moment rotates in response to an applied strain. Dipole coupling 
	drags the bead (rotor) around the stator. b) Orientation of the 
	magnetic moment and principal strains. c) Dipolar forces on a bead near the 
	stator 
	element. }
\end{figure*}

Figure \ref{fig:dipole_motor} shows the a) dipole field, b) stator geometry, 
and c) dipole forces for the motor being analyzed. This figure is a top 
down view of the motor shown in Figure \ref{fig:motor power density scaling}b.
A circular thin-film stator is studied, with in-plane magnetic orientation $ 
\theta $. The stator is subject to a voltage induced plane strain ($\ee_{13} = 
\ee_{23} = \ee_{33} = 0 $), with principal strains components $\ee_1$ and 
$\ee_2 $ rotated $\varphi_{\ee} $ from the x-axis ([100] direction). While the 
magnitude and sign of the applied strain may change as a function of time, the 
orientation is assumed fixed during motor operation (i.e., applied using a 
single fixed electrode). Further assuming the out-of-plane magnetization $ m_3 
$ is small due to shape anisotropy, the relevant stator energy density reduces 
to
\begin{align}
	U_{tot} &= U_{MCA} + U_{ME} \label{eq:Utot}\\
	&= K_1^{MCA} \left[ m_1^2m_2^2 \right] + B_1 	
	\left[\ee_{11}(m_1^2-\nicefrac{1}{3})  + \ee_{22}(m_2^2-\nicefrac{1}{3}) 
	\right] + 2 B_2 \left[\ee_{12}m_1m_2 \right] \label{eq:thin film energy}
\end{align}
where  $ U_{tot} $, $ U_{MCA} $, and $ U_{ME} $ are the total, 
magnetocrystalline, and magnetoelastic energies, respectively, $ K_1^{mca} $ is 
the first order cubic MCA constant, $ B_1 $ and $ B_2 $ are the cubic 
magnetoelastic anisotropy coefficients, the $ m_i $ terms are the magnetization 
direction cosines, and $ \ee_{ij} $ are the applied strain components. Note 
that a factor of 2 has been included on the $ \mathrm{B_2} $ term, as tensorial 
strain components have been used in place of the engineering strain components 
commonly used in this expression (i.e., $ e_{ij}=2\ee_{ij} $ for $ i \neq j 
$).\cite{Chikazumi2009a, OHandley1999} The tensorial strain components are now 
written in terms of the principal strains using a rank 2 tensor transformation
\begin{align}
	\ee_{11} &= 
	\ee_{avg}+\frac{1}{2}\ee_b\cos(2\varphi_{\ee})\label{eq:e_11}\\
	\ee_{22} &= 
	\ee_{avg}-\frac{1}{2}\ee_b\cos(2\varphi_{\ee})\label{eq:e_22}\\
	\ee_{12} &= \frac{1}{2}\ee_b \sin(2 \varphi_{\ee}) \label{eq:e_12}
\end{align} 
where $ \varphi_{\ee} $ is the principal strain orientation shown on Figure 
\ref{fig:dipole_motor}b, $\ee_{avg}=(\ee_1 + \ee_2) / 2 $ is the 
average strain, and $ \ee_b = \ee_1 - \ee_2 $ is the biaxial strain . 

Using equations \ref{eq:e_11} to \ref{eq:e_12} in equation \ref{eq:thin film 
energy} and converting to polar coordinates ($ m_1=\cos\theta $, and $ 
m_2=\sin\theta $), the energy expression becomes
\begin{align}
	U_{tot} &= K_2 \cos(2\theta-\delta_s) + K_4 \cos(4\theta) \label{eq:Utot 
	simplified} \\
	K_2 &=	\frac{1}{2}\ee_b\sqrt{B_1^2\cos^2(2\varphi_{\ee}) 
		+B_2^2\sin^2(2\varphi_{\ee})} \label{eq:K2}\\
	K_4 &= -\frac{1}{8}K_1^{MCA} \label{eq:K4}\\
	\tan(\delta_s) &= \frac{B_2}{B_1} \tan(2 \varphi_{\ee}) 
	\label{eq:delta_s}
\end{align}
where $ K_2 $ and $ K_4 $ are the total second and fourth order anisotropy 
coefficients, and $ \delta_s $ is the static rotation of the second order 
anisotropy with respect to the x-direction. Equation \ref{eq:Utot simplified} 
is indicative of the fact that a conservative anisotropy energy can be 
expanded in a Fourier series ($ U_{tot}= Re[\sum\limits_{n}K_n \Exp{\jj n 
\theta}])$, and highlights the two and four fold rotational symmetry 
inherent to the cubic magnetoelastic and magnetocrystalline anisotropies, 
respectively. It should be noted that in these equations the second order 
anisotropy $ K_2\rightarrow K_2(t) $ is a time dependent function of the 
voltage induced biaxial strain $ \ee_b $, and independent of the 
average strain $ \ee_{avg} $. For an isotropic magnetoelastic material 
($\lambda_{100}= \lambda_{111} \implies  B_1=B_2 $) these expressions 
simplify to $ K_2= \ee_b \left| B_1 \right|/2$ and $\tan(\delta_s)=
\tan(2\varphi_{\ee}) $.

To locate the equilibrium configuration for an arbitrary 
input strain equation \ref{eq:Utot simplified} is first converted to a
complex polynomial 
\begin{align}
	U_{tot} &= Re\left[K_2 \Exp{\jj(2\theta-\delta_s)} + K_4 
		\Exp{4\jj \theta}\right] \\
	&= \frac{1}{2}K_2\left(p z^2 + \overline{pz}\,^2 \right)  + 
		\frac{1}{2} K_4 \left(z^4 + \overline{z}\,^4\right)
\end{align}
where $ p = \Exp{-\jj\delta_s} $ and $ z = \Exp{\jj \theta}$. The extrema of 
this expression are found using the Lagrange multiplier method to enforce 
the constraint that $ z $ has unit magnitude.
\begin{align}	
	\mathcal{L}(z,\conj{z},\lambda) &= U_{tot}(z,\conj{z}) - \lambda(z 
	\conj{z}-1) \label{eq:L} 
\end{align}
Combing the partial derivatives of equation \ref{eq:L} and making the 
substitution $ R = K_2/K_4 $, yields the $ 8^{th}  $ order complex polynomial
\begin{align}
	z^8 + \frac{1}{2}R p z^6 - \frac{1}{2}R \conj{p} z^2 -1=0
\end{align}
which can be converted into a $ 4^{th} $ order polynomial with the substitution 
$ w = z^2 $
\begin{align}
	w^4 + \frac{1}{2}R p w^3 - \frac{1}{2}R \conj{p} w -1=0 \label{eq:w}
\end{align}		
The roots of this equation provide the local extrema of the energy landscape. 

Noting that \emph{R} is the ratio of the second and fourth order anisotropies, 
it is easy to see when no strain is applied $( R=0 )$ energy extrema are 
located 
at the $ 8^{\mathrm{th}} $ roots of unity $( w^4=1\implies 
z^8=1\implies\varphi = \pm n \pi/4)$. Since the roots provide the location 
of both the maxima and minima of the fourth order cubic magnetocrystalline 
anisotropy, the solution for \emph{z} has 8 angles per $ 2\pi $ radians when $ 
R=0 $. The closed form solution for an arbitrary plane strain state is provided 
in the Supplementary Information.


\subsection{Dynamic Analysis}
This section analyzes the dynamics of the proposed motor using the 
Landau-Lifshitz-Gilbert (LLG) equation. The LLG equations is
\begin{align}
	\dot{\vec{m}} &= -\gamma \vec{m} \times \vec{H}_{eff} - \alpha \vec{m} 
	\times \dot{\vec{m}} \label{eq:LLG}	\\
	\vec{H}_{eff} &= -\frac{1}{\mu_0 M_s} \frac{\partial U_{tot}}{\partial 
	\vec{m}} \label{eq:Heff}	
\end{align}
where the gyromagnetic ratio $ \gamma $, Gilbert damping factor $ \alpha $, 
saturation magnetization $ M_s $, and effective magnetic field $ \vec{H}_{eff} 
$ have been used. The effective field has one component for each of the energy 
expressions in Equation \ref{eq:Utot}, with components
\begin{align}	
	H_1 &= -\frac{2}{\mu_0 M_s} \left[K_1^{mca} m_1 m_2^2 + B_1 m_1 
		\ee_{11} + B_2 m_2 \ee_{12} \right] \label{eq:H1}\\
	H_2 &= -\frac{2}{\mu_0 M_s} \left[K_1^{mca} m_1^2 m_2 + B_1 m_2 
		\ee_{22} + B_2 m_1 \ee_{12} \right] \label{eq:H2}\\
	H_3 &= -\frac{2}{\mu_0 M_s} \left[K_1^{mca} m_3( m_1^2+ m_2^2) + K_2^{mca} 
	m_1^2 m_2^2 
	m_3 + \frac{1}{2}\mu_0 M_s^2 m_3 \right] \label{eq:H3}
\end{align}
where fields linear in $ m_3 $ have been retained, but all higher order 
contributions dropped. 

The following analysis determines the strain amplitude and frequency required 
to cause rotational motion with constant angular velocity. It is assumed $ 
m_1(t)=\cos\omega_0 t $, $ m_2(t)=\sin\omega_0 t $, and that the biaxial strain 
follows a temporal dependence of $ \ee_b\propto \cos(\omega_{\varepsilon} t + 
\delta_d) $, where $\omega_{\ee}$ is the mechanical driving frequency, and $ 
\delta_d $ is dynamic phase difference between $ \ee_b $ and $ m_i $. It is 
also assumed that the mechanical strain is applied at twice the frequency of 
rotation $ \omega_{\ee} = 2\omega_0 $, with one strain cycle causing a maximum 
of $ 180^{\circ} $ rotation. This assumption will be justified after looking at 
the results of the quasi-static analysis.

Using the fact that the magnitude of the magnetic moment is conserved 
\begin{align}
	\vec{m}\cdot\dot{\vec{m}}&\approx m_1 \dot{m}_1 + m_2 \dot{m}_2 =0
\end{align}
where $ \dot{\vec{m}} $ is calculated by inserting Equations \ref{eq:H1} to 
\ref{eq:H3} into Equation \ref{eq:LLG}. This results in an expression of the 
form
\begin{align}
	A \sin(4\omega_{\ee} t)+B \cos(4\omega_{\ee} t) +C = 0 
\end{align}
where A, B, and C are functions of the material properties and operating 
conditions but independent of time. For the equality to hold for all time t, 
the coefficients A, B, and C must be identically zero, resulting in the 
operating conditions to achieve uniform rotation. 
\begin{align}
	\omega_0 &= \frac{\gamma K_1^{mca} B_1 B_2 }{\alpha \mu_0 M_s(B_1^2 
		\cot(2\varphi_{\ee}) +B_2^2 \tan(2 \varphi_{\ee}))} 
		\label{eq:omega0}\\
	\ee_b &= \pm\frac{K_1^{mca}}{B_1}	
	\frac{\sec(2\varphi_{\ee})}{\sqrt{1+\nicefrac{B_2^2}{B_1^2}\tan^2(2 
	\varphi_{\ee})}} 
		\label{eq:e_b}\\
	\tan\delta_d &= \frac{B_2}{B_1} \tan(2\varphi_{\ee}) \label{eq:tand}
\end{align}
From equations \ref{eq:omega0} and \ref{eq:e_b}, it should be noted that the 
strain amplitude and frequency required for constant precession depend on both
the  material's anisotropy coefficients and orientation of the the principal 
strain. The dynamic phase difference between the strain and magnetic moment in 
equation \ref{eq:tand} is equal to the static rotational offset of the 
magnetoelastic anisotropy provided in equation \ref{eq:delta_s} (i.e. 
$\tan\delta_d=\tan\delta_s $). While equations \ref{eq:omega0} - \ref{eq:tand} 
provide the conditions for constant angular velocity, applying a larger strain 
than  $ \ee_b $ will produce faster rotation, albeit with non-constant angular 
velocity. For isotropic magnetoelastic materials, these equations simplify to 
\begin{align}
	\omega_0^{iso} &= \frac{\gamma K_1^{mca} }{2\alpha \mu_0 M_s} 
		\sin(4\varphi_{\ee}) \label{eq:omega0iso}\\
	\ee_b^{iso} &= \pm\frac{K_1^{mca}}{B_1}\label{eq:e_biso}\\
	\tan\delta_d^{iso} &= \tan(2\varphi_{\ee}) \label{eq:tandiso}
\end{align}
where the resulting frequency $ \omega_0^{iso} $ increases with large 
crytalline anisotropies, and decreases with larger Gilbert damping and 
saturation magnetization. Additionally, the required strain amplitude $ 
\ee_b^{iso} $ depends exclusively on the ratio of the crystalline and 
magnetoelastic anisotropy coefficients. 

In addition to determining the constant precession conditions, the LLG equation 
was numerically simulated using Matlab, with a macrospin model based on the 
dynamic equations of motion. This was done to verify the calculations performed 
in this section, and to study the behavior of the motor at frequencies above 
and below $ \omega_0 $.

\subsection{Motor Analysis}
The analysis outlined in the preceding sections controls the rotational 
characteristics of the stator's magnetic moment. Assuming the uniformly 
magnetized rotor / stator can be treated as point dipoles, the stator's dipole 
field (due to magnetic moment $\mm_1 $) and force on the rotor (magnetic moment 
$ \mm_2 $) due to the rotating moment of the stator is provided by
\begin{align}		
\vec{B}_1 (\mm_1,\vec{r}) &= \frac{\mu_0}{4 \pi r^3} \left( 
 	3 \left( \mm_1\cdot \hat{r} \right)\hat{r} -\mm_1 
 	\right)\\		
\vec{F}(\mm_1,\mm_2,\vec{r}) &=
		\nabla\left(\mm_2 \cdot \vec{B}_1  
		(\mm_1,\vec{r})\right) \\
	\therefore
\vec{F}(\mm_1,\mm_2,\vec{r}) &= \left. \frac{3\mu_0}{4 \pi r^5} 
	\right[
		\left(\mm_1\cdot\vec{r}\right) \mm_2 + 
		\left(\mm_2\cdot\vec{r}\right) \mm_1 +		
	\ldots \nonumber \\
		&\quad \left(\mm_1\cdot\mm_2 \right) \vec{r}- 
		\left. \frac{5 \vec{r} }{r^2} 
		\left(\mm_1\cdot\vec{r}\right) 
		\left(\mm_2\cdot\vec{r}\right) 
	\right]
\end{align}
where $ \mm_{\alpha} = V_{\alpha} \vec{m}_{\alpha} M_s$ is the vector net 
magnetic moment of the rotor / stator, $ V_{\alpha} $ is the volume, $ M_s $ 
the saturation magnetization, and $ \vec{m}_{\alpha} $ is the direction cosine 
unit vector. Figure \ref{fig:dipole_motor}c shows the resulting force vectors 
at several locations around the stator. For a bead with magnetic moment of 
amplitude $\abs{\mm_2}$ located a distance $ r $ from the stator with moment 
amplitude $ \abs{\mm_1} $, the maximum radial and tangential forces are 
given by Equations \ref{eq: Fr_max} and \ref{eq: Fphi_max}. The radial force is 
maximized when $ \vec{r} \cdot \mm_1 = 0$. The maximum tangential force occurs 
when the angle between the radius and $ \mm_1 $ is $ n \pi \pm 
\mathrm{asin}\left(\sqrt{2} \right) $.
\begin{align}
	\abs{\vec{F}_r} &= \frac{3 \mu_0 \left| \mm_1 \right| 
	\left| \mm_2 \right| } {2 \pi r^4} \label{eq: Fr_max}\\
	\abs{\vec{F}_{\varphi}} &=  \abs{\vec{F}_{r}}/6 \label{eq: Fphi_max}
\end{align}
For both of these equations is has been assumed that $ \mm_2 \parallel 
\vec{B}_1 $. 

\section{Results}	\label{sec: Motor Results}
This section presents the quasi-static energy equilibrium states, followed by 
results from the analytic and numerical treatment of the dynamic LLG response. 
Lastly, the relation between the dynamic response and overall motor power 
density will be discussed. 

\subsection{Quasi-Static Energy Analysis Results}
\begin{figure*}[hptb!]
	\includegraphics[width=1.0\linewidth]{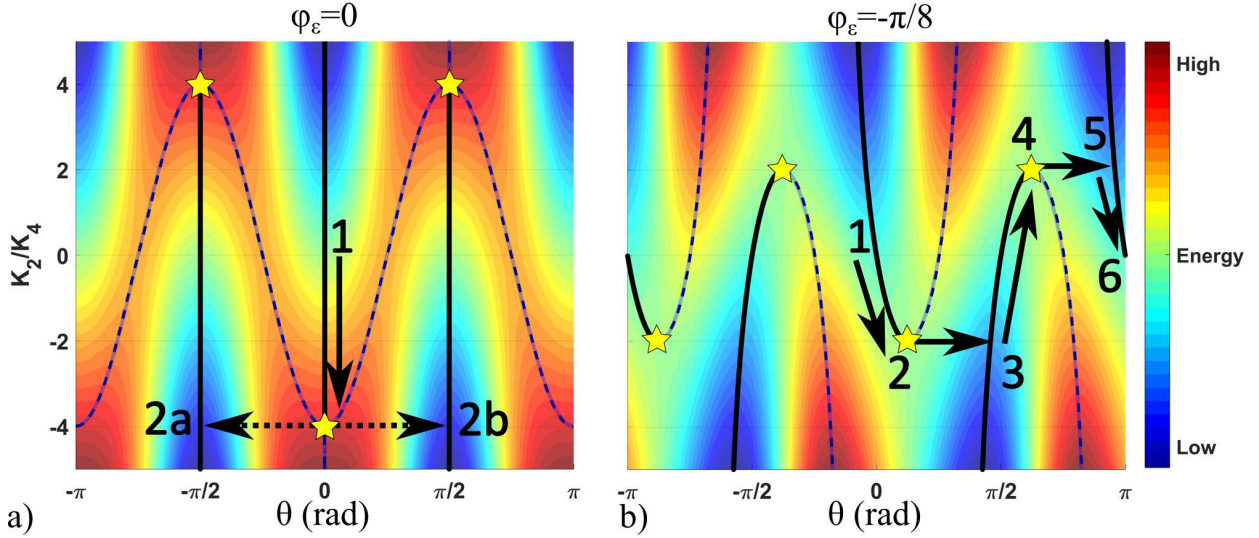}
	\caption{\label{fig:EnergyLandscape} Energy landscapes for two different 
	principal strain orientations ($\varphi_{\ee}$). Solid black lines 
	track energy minima and dotted blue lines track energy maxima. Numbered 
	black arrows correspond to	loading pathways described in the text. a) 
	Non-deterministic switching occurs when the $\varphi_{\ee}=0$. b) 
	Deterministic switching occurs when $\varphi_{\ee}=-\pi/8$ }
\end{figure*}

Figure \ref{fig:EnergyLandscape} shows the energy landscape for two different 
principal strain orientations ($ \varphi_{\ee} $). The x-axis represents the 
orientation of the in-plane magnetic moment $ \theta $, the y-axis shows the 
anisotropy ratio $ K_2/K_4 $. Using Equations \ref{eq:K2} and \ref{eq:K4} 
positive $ K_2/K_4 $ values corresponds to tension and negative to compression. 
Surface color represents the energy for each given state calculated from 
Equation \ref{eq:Utot simplified}. The solid black lines trace the locations of 
energy minima, and dashed blue lines trace the locations of energy 
maxima. Both sets of lines correspond to the roots of the characteristic 
polynomial provided by Equation \ref{eq:w}. Yellow stars indicate inflection 
points in the energy landscape which are locations where the system 
irreversibly flips from one energy minima to another. 

For example, suppose the system in Figure \ref{fig:EnergyLandscape}a starts 
unstrained ($ K_2/K_4=0 $) at position 1, in the energy well at $\theta=0 $ 
radians. If tensile strain is applied (moving up the y-axis), then 
the system moves into a deeper energy well and doesn't rotate. If compressive 
strain is applied, the system moves down the black line until $ K_2/K_4=-4 $, 
at which point the magnetic moment non-deterministically flips to energy wells 
at $\theta = \pm \pi/2 $ corresponding to positions 2a and 2b. 

The case of deterministic switching is shown in Figure 
\ref{fig:EnergyLandscape}b, where $ \varphi_{\ee}=-\pi/8=-22.5^{\circ} $. In 
this case starting unstrained in the energy well at position 1 ($ \theta=0 $), 
compression again moves the system down the y-axis, but also slightly rotates 
it in the counter-clockwise direction until it reaches position 2. When $ 
K_2/K_4=-2 $ an inflection point is encountered and the system 
deterministically flips to position 3, the energy well at $ \theta \approx 
\pi/2 $. Applying tension to the system until $ K_2/K_4=+2 $ causes 
counter-clockwise rotation to the inflection point at position 4, and the 
system then jumps to position 5 ($ \theta \approx \pi$), and removing the 
strain moves it to position 6. This example demonstrates how a single cycle of 
biaxial compression / tension causes the magnetic moment to deterministically 
rotate $ 180^{\circ} $. Therefore, two compression / tension cycles are needed 
to rotate a full $ 360^{\circ} $, and the system rotates at half the driving 
frequency (i.e., $\omega_{\ee}=2\omega_0 $). This indicates that it is possible 
to deterministically rotate a magnetic moment using a magnetoelastic anisotropy 
with a fixed direction (i.e., allowing single electrode control). 

\begin{figure}[hptb!]
 	\centering	
	\includegraphics[width=8.5cm]{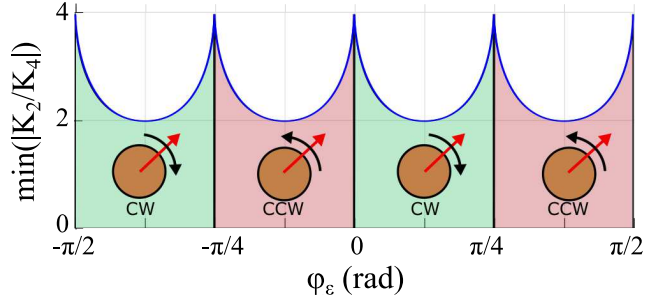}
	\caption{\label{fig:min strain} Minimum anisotropy ratio required for 
	rotation to occur. Deterministic rotation occurs as long as $ 
	\varphi_{\ee} \neq 	n \pi/4 $. The minimum flipping strain occurs when 
	$\varphi_{\ee} = \pi/8 \pm n\pi/4 $} 
\end{figure}

Figure \ref{fig:min strain} shows the minimum anisotropy ratio required to 
rotate the magnetic moment. The upper solid blue line corresponds to the 
minimum ratio $ K_2/K_4 $ required to rotate, while the surface color and 
insets indicate the direction of magnetic rotation. When $
\varphi_{\ee} = \pm n \pi/4 $ the magnetic moment changes rotation directions. 
As $ \varphi_{\ee} =\pm n \pi/4 $ corresponds the $ \langle 1 0 0 \rangle $ and 
$ \langle 1 1 0 \rangle $ family of directions, it becomes clear that 
non-deterministic rotation results when strain is applied along a 
crystallographic direction with large degree of symmetry. When $ \varphi_{\ee} 
= \pi/8 \pm n\pi/4 $, the system is rotated with the minimum possible 
anisotropy ratio ($ \abs{K_2/K_4}=2$). For isotropic magnetoelasticity, the 
minimum biaxial strain required to generate quasi-static rotation is 
\begin{align}
 \min(\abs{\ee_b^{iso}}) = 
  \begin{cases}
     \frac{K_1^{MCA}}{2 B_1}& 
    	\text{for }\varphi_{\ee}=\frac{\pi}{8}\pm \frac{n\pi}{4} \\
    \frac{K_1^{MCA}}{B_1}& 
        	\text{for }\varphi_{\ee}=\pm \frac{n \pi}{4} 
  \end{cases} \label{eq:min eb static}
\end{align}

Applying strain at the appropriate angle reduces the flipping strain amplitude 
by a factor of 2, and therefore lowers the strain energy by a factor of 4 
compared to the non-deterministic case. Based on Figure \ref{fig:min strain}, a 
two input system generating principal strain states oriented at $ \pm \pi/8 $ 
will control bipolar operation. Lastly, while the fourfold anisotropy has been 
assumed to be crystalline in nature, shape anisotropy can also produce cubic 
anisotropies, as experimentally demonstrated by \textcite{Lambson2013}. 

\subsection{Dynamic Results}
Recall that the biaxial strain and frequency required to obtain constant 
precession were previously presented in Equations \ref{eq:omega0}-\ref{eq:tand} 
for a general material, and Equations \ref{eq:omega0iso}-\ref{eq:tandiso} for 
an isotropic material. For isotropic magnetoelasticity, the strain amplitude 
required for rotation at constant angular velocity matches the strain required 
for non-deterministic flipping to occur in the quasi-static analysis ($ 
\ee_b^{iso} = \pm \frac{K_1}{B_1} $). Furthermore, the rotational frequency 
depends on the orientation of the applied strain ($ \omega_0^{iso} 
\propto \sin(4\varphi_{\ee}) $). When the strain is applied at an angle $ 
\varphi_{\ee}=\pm n \pi/4 $, the angular velocity $ \omega_0=0 $. Therefore, 
when $ \varphi_{\ee}=\pm n \pi/4 $, both the quasi-static and dynamic analyses 
converge to predicting $ \ee_b^{iso} = \pm \frac{K_1}{B_1} $, with zero angular 
frequency (i.e., random non-deterministic rotations). This demonstrates 
consistency between the static and dynamic solutions. Consistency was further 
checked by numerically simulating the motor stator element operating at several 
strain amplitudes and frequencies.  

\begin{figure*}[hptb!]
	\centering		
	\includegraphics[width=1.0\linewidth]{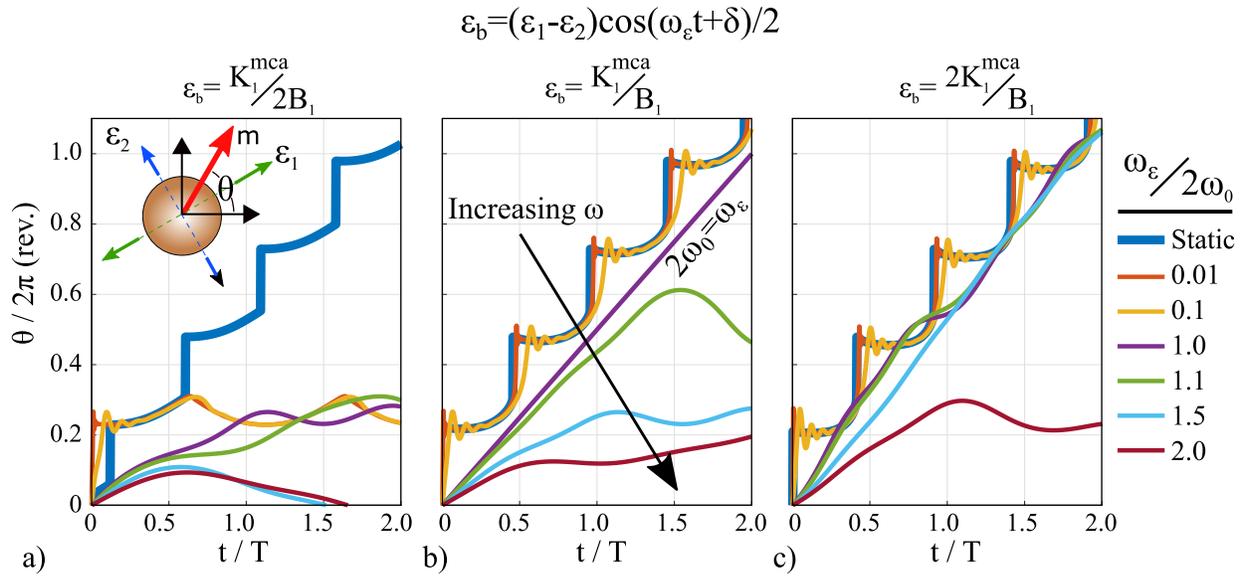}
	\caption{\label{fig: rev vs time} Magnetic revolutions vs time for a 	
	motor with isotropic magnetoelasticity and $ \varphi_{\ee}=-\pi/8$ 
	at several 	driving frequencies. Results are shown temporally normalized 
	with respect to the period of the strain signal ($T=2\pi / 
	\omega_{\varepsilon}$). Simulations are shown for Gilbert damping $ 
	\alpha=0.01 $. Specific	material properties and relevant model terms are 
	presented in Table 	\ref{table:motor materials}. } 
\end{figure*}

Figure \ref{fig: rev vs time} shows the frequency response of this motor stator 
for quasi static and dynamic operation at 7 different frequencies for two full 
cycles of tension / compression. Parts a)-c)  show the impact of using three
different biaxial strain amplitudes. At quasi-static rates (i.e., calculated 
tracking energy equilibrium orientations), the magnetic moment exhibits 
ratchet-like behavior, with abrupt switches every time the minimum 
switching strain is reached ($ \ee_b = \pm\frac{K_1^{MCA}}{2 B_1} $). If a 
larger strain is applied (shown in part b and c) the magnet undergoes a small 
period of clockwise motion (i.e., temporarily changes direction), as would be 
expected from Figure \ref{fig:EnergyLandscape}. 

Figure \ref{fig: rev vs time}a indicates that the minimum predicted strain 
generates continuous rotation only for quasi-static behavior. Operating 
at $\omega_{\varepsilon}/ 2\omega_0 \geq0.01$ when $ \ee_b = \pm 
\frac{K_1^{MCA}}{2B_1} $ results in minor fluctuations about a fixed 
orientation. While not depicted in Figure \ref{fig: rev vs time}a, it was 
observed that increasing the strain to $\varepsilon_b =1.01 \cdot 
\frac{K_1^{MCA}}{2B_1} $ generated successful rotation for 
$\omega_{\varepsilon}/ 2\omega_0\leq0.01 $, indicating the minimum flipping 
strain is valid at frequencies that are small compared to $ \omega_0 $. The 
under damped yellow line in Figure \ref{fig: rev vs time}b indicates that if $ 
\ee_b = \pm\frac{K_1^{MCA}}{B_1} $, quasi-static performance is maintained up 
to $ \omega_{\varepsilon}/ 2\omega_0 \leq 0.1 $. Furthermore, the case where $ 
\ee_b = \pm\frac{K_1^{MCA}} {B_1} $ and $\omega_{\varepsilon} =2\omega_0 $ 
resulted in constant angular velocity as analytically predicted by Equation 
\ref{eq:omega0iso} (shown with the straight purple line). If the frequency $ 
\omega_{\varepsilon} $ is further increased by $ 10\%$, the motor fails to 
deterministically rotate. 

Figure \ref{fig: rev vs time}c shows that when larger strains are 
applied, larger operational frequencies can be attained. Doubling the strain 
from part b results in constant precession with a $ 50\% $ increase in 
rotational frequency. It should be noted that this increase is not directly 
proportional, as operating with double the strain amplitude did not enable 
operation at twice the frequency. The data in Figure \ref{fig: rev vs time} was 
reproduced for a variety of Gilbert damping coefficients from 0.01 to 0.1, with 
Figure \ref{fig: rev vs time} showing results for $ \alpha=0.01 $. They were 
found to provide qualitatively similar results, albeit with the highest damping 
leading to smaller overshoot and smoother overall behavior. 

Table \ref{table:motor materials} provides a list of magnetic materials along 
with their relevant properties, predicted minimum strain for quasi-static 
operation, and operational frequency for constant precession. This 
table makes it abundantly clear that the ratio $ K_1^{mca}/B_1 $ is the key 
metric to reduce the applied strain amplitude. Moderately magnetoelastic 
materials like Ni and $\mathrm{Ni_{55}Fe_{45}}$ require the smallest 
strains ($90\leq\ee_b\leq360$ppm) due to their low crystalline anisotropies. On 
the other hand, an epitaxially grown thin film of Terfenol-D 
($\mathrm{Tb_{0.3}Dy_{0.7}Fe_{2}}$) has the largest magnetoelastic anisotropy 
coefficient $ B_1 $, but due to a large crystalline anisotropy $ K_1 $ would 
require $\ee_b\approx3,300$ ppm to rotate, which would be challenging for most 
ferroelectric materials even at quasi-static frequencies. 

\begin{table*}[htpb]
\caption{ \label{table:motor materials}
Comparison of the minimum strain required for deterministic control of a 
variety of magnetoelastic materials.}
\begin{ruledtabular}
\begin{tabular}{lcccccc}
\textrm{} &
\textrm{$ \mathrm{\mu_0 M_s} $} &
\textrm{$ \mathrm{K_1} $} &
\textrm{$ \mathrm{B_1}$
	\footnote{$\mathrm{B_1=-\frac{3}{2}\lambda_{100}(C_{11}-C_{12})} $} 
	\footnote{For isotropic materials $\mathrm{(C_{11}-C_{12})=E/(1+2\nu) }$}} &
\textrm{$\mathrm{\varepsilon_b^{min}=\frac{K_1}{2B_1}}$} & 
\textrm{$\mathrm{\omega_0 / 2\pi=\frac{\gamma K_1^{mca} }{4 \pi \alpha \mu_0 
M_s} }$}
 	\footnote{Assumed $\mathrm{\alpha=0.01} $ for all materials}  
& 
\textrm{} \\
\textrm{Material} &
\textrm{(\SIlist{}{\tesla})} &
\textrm{(\SIlist{}{\kilo\joule\per\meter^3})} &
\textrm{(\SIlist{}{\mega\joule\per\meter^3})} &
\textrm{(ppm)} &
\textrm{(GHz)} &
\textrm{Source} \\
\colrule
Ni-FCC & 0.6 & -4.5 & 6.2 & 363 & 13 &\cite{OHandley1999} \\
Co-FCC & 1.8 & -120 & -16 & 3,750 & 117 &\cite{OHandley1999} \\
Co-HCP & 1.8 & 350 & 6 & 29,167 & 342 &\cite{OHandley1999} \\
Fe-BCC & 2.1 & 48 & -2.9 & 8,276 & 40 &\cite{OHandley1999}\\
$\mathrm{Fe_{81}Ga_{19}} $\footnote{ Assuming E=75 GPa, $ \mathrm{\nu=0.3 }$} 
	& 1.4 & 17.5 & 22.4 & 391 & 22 &\cite{Atulasimha2011} \\
$\mathrm{Tb_{0.3}Dy_{0.7}Fe_{2}}$\footnote{Epitaxial thin film}  
	& 0.9 & -525 & -80 & 3,280 & 1,027 &\cite{Fuente2004} \\
$\mathrm{Ni_{55}Fe_{45}}$ & 1.6 & 1 & 5.5 & 90 & 1.1 &\cite{OHandley1999} 
\end{tabular}
\end{ruledtabular}
\end{table*}

The frequencies listed in Table \ref{table:motor materials} range from 
\SIlist{1}{\giga\hertz} to \SIlist{1}{\tera\hertz}, and indicate the magnetic 
dynamics of these motors will have minimal effect once mechanical losses are 
accounted for. Also recall that quasi-static behavior was seen to persist up to 
$ 1\% $ of $ \omega_0 $ in Figure \ref{fig: rev vs time}b. That means these 
motors are magnetically quasi-static at frequencies up to 10s of 
\SIlist{}{\mega\hertz} (i.e., in energy equilibrium), providing broadband 
operation. It should be noted that the Gilbert damping parameter was assumed to 
be $ \alpha = 0.01 $ for all the listed materials, while larger values may be 
encountered. At the frequencies predicted in this paper, the coupled mechanical 
behavior of the motor will be important, and needs to be incorporated for more 
thorough and accurate predictions (i.e., using a fully coupled modeling 
approach), however that is consider outside the scope of the present article. 

\subsection{Multiferroic Motor Power Density}
The last focus of this paper is to determine approximate performance bounds of 
a mechanically lossless bead-on-on-a-disk motor using the developed 
control scheme. The material from Table \ref{table:motor materials} requiring 
the smallest strain is $\mathrm{Ni_{55}Fe_{45}}$, which is the focus of this 
section.  A motor with stator radius $r_1=$\SI{100}{\nano\meter}, thickness 
$t=$\SI{10}{\nano\meter} and rotor (bead) radius of $r_2=$\SI{10}{\nano\meter} 
was simulated, with dimensions chosen to keep the stator a single domain. 

In the absence of mechanical losses, the upper bound for the rotary motor's 
frequency is dictated by inertial forces. In other words, if the rotary motor 
is operated at too large a frequency, the bead will be flung away from the 
stator due to its own momentum. While this may be useful in some circumstances, 
it does limit the maximum frequency to
\begin{align}
	\abs{\vec{F}_{r}} &= m \omega^2 r	\\ 
	\therefore \quad f_{max} &= \frac{1}{2 \pi} \sqrt{\frac{3 \mu_0 
				\left|\mm_1\right| \left|\mm_2\right|}{2 \pi m 
				r^5}}			 
\end{align}
where Equation \ref{eq: Fr_max} has been used. For the specified motor, $ 
f_{max} \approx$\SI{9.7}{\mega\hertz}, which corresponds to a linear 
velocity $ v\approx$\SI{6.1}{\meter\per\second}. Based on the results from 
Table \ref{table:motor materials}, this is roughly 100 times slower than $ 
\omega_0^{iso}/2\pi $. However, this low frequency / velocity is a limit of 
this specific rotary motor design, and not an intrinsic limit for multiferroic 
motors. For instance, a linear motor can be made that overcomes these inertial 
limitations.

\begin{figure}[hptb!]
	\centering
	\includegraphics[width=8.5cm]{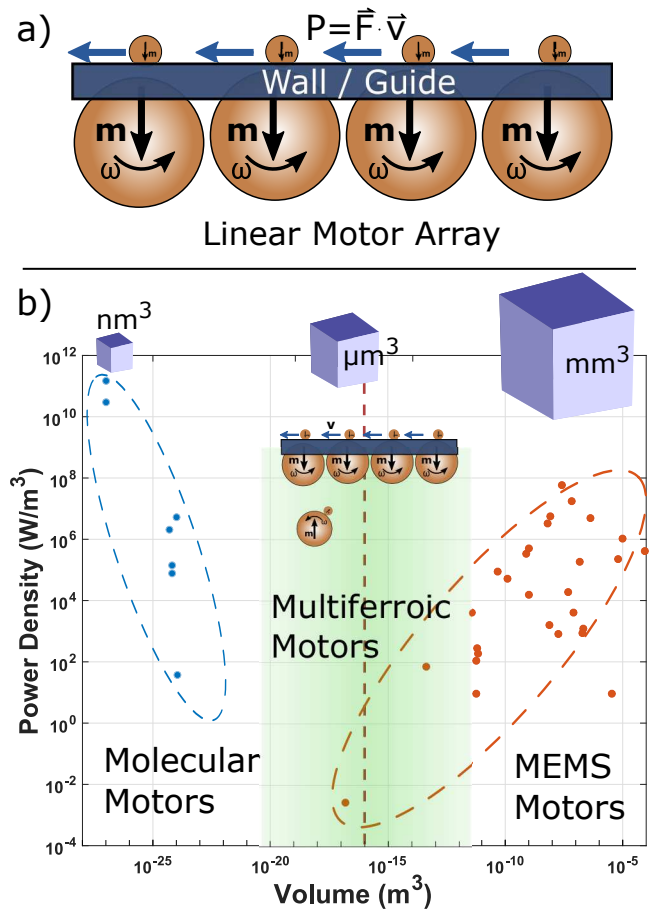}
	\caption{\label{fig:power density updated} Updated power density chart now 
	showing approximate upper bounds for both rotary and linear multiferroic 
		bead-on-a-disk motors.\protect\cite{Balzani2006, Lussis2011,Regan2005, 
		Kang2009, Fuller2007, Junge2009, Paenke2001, Liu2010, Kim2014, Tai1989, 
		Wallrabe1992, Tai1989a, Liu2010a, Jacobsen1989,Ghalichechian2008, 
		Livermore2004, 	Glickman2011, Chang2014, Achotte2006, Takemura2007, 
		Kanda2006, Watson2009, 	Mashimo2015, Flynn1998, Sashida1993, 
		Morita2000, Cagatay2004, Satonobu2000, 	Suzuki2002,	Watson2009a, 
		DeVolder2011, Buetefisch2002, Konishi2001, 	DeVolder2008, 
		Gebhard1997} }
\end{figure}

The linear motor shown in Figure \ref{fig:power density updated}a overcomes the 
frequency limitations of the rotary motor by not forcing the bead to change 
directions (i.e., it works with the bead's inertia). This motor can be 
fabricated by patterning rotary motors adjacent to each other, and optionally 
etching a trough, or depositing a wall / barrier on top of them to physically 
guide the bead. As inertial effects are now advantageous, and help pass the 
bead from one disc to the next, in the limit of zero fluidic drag the bead can 
be propelled as fast as the stator's magnetic moment can rotate.
For a $\mathrm{Ni_{55}Fe_{45}}$ motor with the dimensions listed above, the 
magnetic moment is able to precess at a frequency of $f\approx$ 
\SI{1.1}{\giga\hertz}, which corresponds to a linear velocity of $ 
v\approx$\SI{700}{\meter\per\second}. 

Based on the attainable speed, and dipole 
forces driving the motion, approximate bounds on the maximum power density of  
the rotary and linear motors can be calculated. Assuming the vibrating 
piezoelectric substrate is \SIlist{100}{\micro\meter} thick, the rotary motor 
volume is on the order of $ 10^{-18} \mathrm{m}^3 $. The maximum rotary power 
density is predicted to be \SIlist{4.4}{\mega\watt\per\meter^3}, and the 
maximum linear power density is predicted to be over two orders of magnitude 
higher at  \SIlist{470}{\mega\watt\per\meter^3}. The increase in power density 
is due entirely to the higher operational frequency of the linear motor. These 
predictions provide approximate upper bounds for multiferroic bead-on-a-disc 
motors with negligible mechanical losses.

Figure \ref{fig:power density updated} shows the power density chart now 
updated with predictions for the rotary and linear multiferroic motors. The 
multiferroic motor concept fills the void left by previous technologies, 
providing a power dense motor with \si{\micro\meter\cubed} dimensions. The 
upper limit to this boundary corresponds to motors operating at GHz 
frequencies, with lower power densities achieved by decreasing the operational 
frequency. Furthermore, the linear motor is able to scale up in size while 
still maintaining the same power density. This is accomplished in practice by 
creating particle conveyor belts with the linear motor that have relatively 
large lengths. 

To close, we comment on a key simplifying assumption used in this analysis, 
namely the absence of mechanical losses including friction, stiction, and 
fluidic drag. Operation anywhere close to these large frequencies 
will involve mechanical losses, which need to be a key focus of future studies. 
This analysis highlights the key fact that the intrinsic magnetic dynamics are 
sufficiently fast that the stator element is likely to always be in a state of 
energy equilibrium during actual use. This suggests two very important motor
characteristics. One, since stator speed is not a fundamental limit of these 
devices, future researchers would be prudent to focus on designs moving 
magnetic structures like domain walls and onion states with large magnetic 
gradients that amplify dipolar forces. Such a structure has been analyzed by 
\textcite{Sohn2015}, who dragged a \SIlist{1}{\micro\meter} diameter 
magnetic bead at speeds near \SIlist{1}{\milli\meter\per\second}. That motion 
likely achieved a power density on the order of 
\SIlist{10}{\kilo\watt\per\meter^3}, a point right in the middle of the power 
densities predicted in Figure \ref{fig:power density updated}. Second, the 
expected quasi-static behavior of the stator also implies it's rotational 
behavior will be similar to a stepper motor, with abrupt transitions. This 
characteristic would complicate the use of a rotary motor design, with rapid 
changes shaking the beads loose. However a linear motor should still work even 
with abrupt transitions, as the bead's own momentum becomes an aid to continued 
motion. 

\section{Conclusion} \label{sec: Motor Conclusion}
This work has demonstrated how controlling the relative orientation of a 
two-fold anisotropy (magnetoelasticity) with a four-fold anisotropy (e.g., 
cubic MCA) can achieve single electrode control of deterministic 
360 degree magnetic rotations. Bounds were determined for the strain 
required to generate continuous rotation for quasi-static and dynamic 
operation, and the conditions for achieving constant angular precession were 
determined. For currently available materials, strains as low as 90 ppm are 
predicted to enable the operation of these motors, and large frequencies are 
attainable in the absence of fluidic damping. A novel linear motor was also 
proposed, that avoids some limitations of the rotary bead-on-a-disk motor. The 
proposed motors are predicted to have large power densities at a size scale 
where no power dense alternative technologies currently exist, and 
offer a promising new technology for further exploration.


\begin{acknowledgments}
This research was supported by the NSF Nanosystems Engineering Research Center 
for Translational Applications of Nanoscale Multiferroic Systems (TANMS) 
Cooperative Agreement Award (No. EEC-1160504).
\end{acknowledgments}

\bibliography{refs}

\section{Supplementary Information}

\subsection{Proof Ellipsoidal Magnets Can Not Enable $360^{\circ}$ Deterministic
Single Input Control}
Several previous modeling and experimental studies have focused on controlling 
magnetic structures with elliptical shapes. However, the use of ellipsoidal 
structures alone can not result in deterministic $ 360^{\circ} $ rotations if 
the principal strain axis is stationary (i.e., for single input control). While 
it should be clear from physical arguments that combining magnetoelasticity and 
a uniaxial shape anisotropy (i.e., using an ellipse) is limited to $ 
\leq90^{\circ} $ rotations, we provide the following proof for completeness. 

Assuming without loss of generality that the ellipse has a long axis in the 
y-direction, and that the magnetization is constrained to lie in the xy plane, 
the demagnetization energy is
\begin{align}
	U_{demag} &= K_u \cos(2\theta)
\end{align}
where $K_u = -(N_y-N_x)/2 $ is the uniaxial demagnetization energy coefficient 
which depends on the x and y values of the demagnetization tensor $N_x$ and 
$N_y$. Combining this energy with the arguments that lead to Equation 
\ref{eq:Utot simplified}, magnetoelastic rotations in an elliptical shape are 
governed by the total energy expression.
\begin{align}
	U_{tot} &= K_u \cos(2\theta)+ 
		\frac{1}{2}\ee_b B_1\cos(2\varphi_{\ee})\cos(2\theta) + 			
		\frac{1}{2}\ee_b B_2\sin(2\varphi_{\ee}) \sin(2\theta) 
		\label{eq: U with shape}
\end{align}
with $ \varphi_{\ee} $ still the principal strain orientation. 

Equation \ref{eq: U with shape} is simply the addition of $ \sin $  and $ \cos 
$ terms of the same frequency. Of course, $ A\cos(n\theta) + B\sin(n\theta) = 
C\cos(n\theta-\delta) $, where $ C = sgn(A)\sqrt{A^2+B^2} $, and $ \delta = 
\tan(B/A)$. Therefore the energy can be written as a single uniaxial anisotropy 
\begin{align}
	U_{tot} &= K_{tot}\cos(2\theta  - \delta) \\
	K_{tot} &= 	
		sgn\left(\frac{1}{2}\ee_b B_1\cos(2\varphi_{\ee})+K_u\right) \ldots 
			\nonumber \\ 
		&\qquad \sqrt{	
		\left(\frac{1}{2}\ee_b B_1\cos(2\varphi_{\ee}) +K_u\right)^2 
		+\left(\frac{1}{2}\ee_b B_2\sin(2\varphi_{\ee})\right)^2
		} \\
	\tan(\delta) &= 		 
		\frac{\ee_b B_2 \sin(2\varphi_{\ee})/2 }{K_u + \ee_b B_1 
	\cos(2\varphi_{\ee})/2}
\end{align}
This energy has local extrema when $ \theta  =  \pm n \pi/4 +\delta/2$, which 
is controlled by $ \delta $,, while the sign of $ K_{tot} $ determines whether 
the locations is a maxima or minima. The equation for $ \tan(\delta) $ can be 
simplified to
\begin{align}
	\tan(\delta) &= \frac{B_2 \sin(2\varphi_{\ee}) }{2 K_u/\ee_b  +  B_1 
	\cos(2\varphi_{\ee})}
\end{align}
In this equation, the only dynamic variable is the biaxial strain $ \ee_b $. 
Examining $ \tan(\delta) $ as $ \ee_b $ approaches several limiting values 
reveals
\begin{align}	
	\lim\limits_{\ee_b\rightarrow 0} \left(\tan(\delta)\right) &= 0\\
	\lim\limits_{\ee_b\rightarrow \pm\infty} \left(\tan(\delta)\right) &= 
		\frac{B_2}{B_1}\tan(2\varphi_{\ee})
\end{align}
The changes in the total anisotropy orientation are constrained to $0 \leq 
\left| \tan(\delta) \right| \leq \left| B_2/B_1 \tan(2\varphi_{\ee})\right| $, 
with the sign controlled by the material properties and orientation of the 
principal strains. While the total anisotropy has the same orientation for 
large $ \pm \varepsilon_b $, the sign of $ K_{tot} $ will change. In one case $ 
\delta $ specifies an energy minimum, while in the other $ \delta $ specifies 
an energy maximum, with minimum rotated $ 90^{\circ} $. Therefore, a maximum
repeatable rotation of $ 90^{\circ} $ can be achieved by combining an 
elliptical element with a magnetoelastic anisotropy where the principal strain 
orientation is constant.

\subsection{Static Energy Analysis: Complex Roots}

The roots of equation \ref{eq:w} were determined using Mathematica, and are 
provided with the closed form expression:
\begin{align}
	w_i(s_1,s_2,s_3) &= -\frac{A}{4}+s_1 \frac{1}{2}\sqrt{\frac{A^2}{4}+d}+ 
	\ldots \nonumber \\ 
	&\quad s_2\frac{1}{2} \left[\frac{A^2}{2} + 			
		\left(\frac{2}{3c}\right)^{1/3} a 
		- \left(\frac{c}{18}\right)^{1/3} + s_3 
		\frac{A^3+8B}{4\sqrt{\frac{A^2}{4} +d} }
		\right] \label{eq:polynomial roots first}\\
	A &= \frac{R p}{2}\\
	B &= -\frac{R \conj{p}}{2}\\
	a &= 4 + A B\\
	b &= A^2 - B^2\\
	c &= -9 b + \sqrt{12 a^3 + 81 b^2}\\
	d &= \left(\frac{1}{36 c}\right)^{1/3} (-2 3^{1/3}a + (2 c^2)^{1/3}) 
	\label{eq:polynomial roots last}
\end{align}
where the $ s_i $ represent sign changes from one root to the next. The roots 
are provided by 
\begin{align}
	w_1 &= w_i(-1,-1,+1)\\
	w_2 &= w_i(-1,+1,+1)\\
	w_3 &= w_i(+1,-1,-1)\\
	w_4 &= w_i(+1,+1,-1)\\
	z_i &= \pm \sqrt{w_i}
\end{align}

\end{document}